# Crystal growth of copper-rich ytterbium compounds: The predicted giant unit cell structures $YbCu_{4.4}$ and $YbCu_{4.25}$

Saskia Gottlieb-Schönmeyer[1,*], Stefan Brühne[1], Franz Ritter[1], Wolf Assmus[1],
Sergiy Balanetskyy[2], Michael Feuerbacher[2],
Thomas Weber[3], Walter Steurer[3]

[1] Johann Wolfgang Goethe-Universität, Max-von-Laue-Str. 1, D-60438 Frankfurt am Main, Germany

[2] Institut für Mikrostrukturforschung, Forschungszentrum Jülich GmbH, D-52425 Jülich, Germany

[3] Laboratory of Crystallography, Department of Materials, HCI G 511, Wolfgang-Pauli-Strasse 10, CH-8093 Zurich, Switzerland

Two new phases $YbCu_{4.4}$ and $YbCu_{4.25}$ are found as a result of careful phase diagram investigations. Between the congruent and peritectic formation of $YbCu_{4.5}$ and $YbCu_{3.5}$, respectively, the phases $YbCu_{4.4}$ and $YbCu_{4.25}$ are formed peritectically at 934(2)°C and 931(3)°C.
Crystal growth was realised using a Bridgman technique and single crystalline grains of about 50-100μm were analyzed by electron diffraction and single crystal X-ray diffraction. Due to the only slight differences in both compositions and formation temperatures the growth of larger single crystals of a defined superstructure is challenging.
The compounds $YbCu_{4.4}$ and $YbCu_{4.25}$ fit in Černý`s (J. Solid State Chem. **174** (2003) 125) building principle $\{(RECu_5)_n \cdot (RECu_2)\}$ where $RE$ = Yb with $n$ = 4 and 3. $YbCu_{4.4}$ and $YbCu_{4.25}$ base on $AuBe_5/MgCu_2$-type substructures and contain approximately 4570 and 2780 atoms per unit cell. The new phases close the gap in the series of known copper-rich rare earth compounds for $n$ = 1, 2 ($DyCu_{3.5}$, $DyCu_{4.0}$) and $n$ = 5 ($YbCu_{4.5}$, $DyCu_{4.5}$).

Keywords: A. rare-earth intermetallics, B. phase diagrams, C. crystal growth, F. diffraction, F. calorimetry

Introduction

As many physical properties of condensed matter depend on translational symmetry, they are ruled by crystallographic symmetry and unit cell dimension. In conventional metals and intermetallics, periodicities of less than 1 nm usually occur and the corresponding unit cells host a few atoms only. On the other hand, a number of alloys exist that contain more than several thousands of atoms per unit cell. These can be termed complex metallic alloys (CMAs). In CMAs, the length scale of the lattice periodicity is up to several nanometres.

---

*Corresponding author. Tel.: +49 069 79847260; fax: +49 069 79847270. E-mail address: gottlieb@physik.uni-frankfurt.de (S. Gottlieb-Schönmeyer)

Including the dimension of the hosted short-range substructure, two different competing length scales are defined with appropriate consequences for the physical properties. In order to determine and understand intrinsic materials properties e. g. electrical and thermal conductivity, magnetic properties etc., phase pure and single crystalline material is required. Moreover, the occurrence of complex structures often is accompanied by a complex phase diagram.

In the focus of this work is the Cu-Yb system that shows huge unit cell sizes. The phase diagram has been studied by Subramanian & Laughlin [1] and Iandelli & Palenzona [2]. The existence of five phases ($YbCu_{6.5}$/$YbCu_5$, $YbCu_{4.5}$, $YbCu_{3.5}$, $YbCu_2$ and $YbCu$) has been reported. But especially the copper-rich region arouses a lot of interest in the recent years [3,4,5]. Latest phase diagram investigations indicated a phase width for the congruently melting phase $YbCu_{4.5}$ [5].

The crystal structure of $YbCu_{4.5}$ was solved in 1996 by Černý et al. [3]. It is a monoclinically distorted 7×7×6.5 superstructure of the cubic $AuBe_5$ structure type with 7448 atoms per unit cell and cell parameters of about 5nm ($a$ = 48.96Å, $b$ = 48.99Å, $c$ = 45.64Å, $\beta$ = 91.2°). It was also verified that the phase $DyCu_{4.5}$ crystallises in the same structure type with similar lattice parameters [4]. A building principle for a series of superstructures was proposed which was confirmed for the phases $DyCu_x$ ($x$ = 3.5, 4.0, 4.5) and $YbCu_{4.5}$. All superstructures are built of two kinds of structural blocks: $AuBe_5$-type and $MgCu_2$-type blocks are combined to comply with the formula $\{(AB_5)_n \cdot (AB_2)\}$, $n$ = 1,2,3,4,5. The relation between $x$ and $n$ is:

$$x = (5n+2)/(n+1) \quad (1)\,[4].$$

It is noteworthy that the phases $AB_{4.4}$ and $AB_{4.25}$ ($n$ = 4 and $n$ = 3) have been predicted according to this building principle, but their existence has not been proved so far. Recently, it was suggested that the phase $YbCu_{3.5}$ follows the same building principle [5]. The assumption that other RE-Cu (RE = rare earth) systems besides Dy-Cu and Yb-Cu contain isostructural phases, which we propose to call Černý phases, is probable but unverified up to now. Especially the phases $RE$Cu$_{4.5}$ that are known for $RE$ = Lu, Yb, Tm, Er, Ho, Dy and Gd may adopt similar huge monoclinic superstructures (mC7448). Despite all similarities, differences between those systems are expected. One example is the monoclinic distortion where β differs between DyCu$_{4.5}$ (β = 97.5°) and YbCu$_{4.5}$ (β = 91.24(1)°) [4]. Physical properties for most of these systems are almost not known. Although polycrystalline material of the phase YbCu$_{4.5}$ has been analysed under ambient and high pressure [6,7,8,9] a detailed study of single crystalline samples of a *defined* superstructure has not been done up to the present. The aim of this contribution is the preparation of single crystalline material of YbCu$_x$ (4.0 < $x$ < 4.5). As ytterbium possesses a high vapour pressure, a closed crucible Bridgman technique was used. Careful investigations by DSC differential scanning calorimetry (DSC), single crystal x-ray diffraction (SC-XRD) and selected area electron diffraction (SAED) reveal the existence of two more superstructures in the Yb-Cu system that follow Černý's building principle: YbCu$_{4.4}$ and YbCu$_{4.25}$.

Experimental

For phase diagram investigations a Simultaneous Thermal Analysis (STA) device (STA 409, Netzsch) has been used, which allows simultaneous thermogravimetric analysis (TGA) and differential scanning calorimetry (DSC). The pure elements ytterbium (99.99%, ChemPur) and copper (99.999%, Koch-Light Laboratories Ltd) were sealed in tantalum crucibles (∅ = 8mm, height = 17mm). First the crucibles were heated with 10K/min up to 1000°C, then annealed up to one hour and cooled with the same rate (10K/min). A more careful analysis was performed by a lower heating rate (4K/min or 2K/min) in the range between 750°C and 1000°C.

After the DSC measurements the crucibles were cut lengthwise by spark erosion and the cut surfaces of the samples were polished with abrasive paper (3μm SiC). All samples were analyzed in a SEM (scanning electron microscope, Zeiss DSM 740A combined with electron diffraction X-ray analysis (EDX, Ametek, EDAX Genesis V 4.52; the accuracy is ± 1at%).

To grow single crystals the pure elements were sealed in a tantalum crucible (Ø = 9mm, height = 85mm). The crucible was heated inductively up to 1000 °C and growth was performed under protective gas (Ar, 1.5bar) by lowering the crucible out of the high frequency coil at 3mm/h (Bridgman technique). The crucibles provide a specially defined seed selction zone at the bottom (Fig. 1a). The temperature was checked pyrometrically. After growth the sample was cut by spark erosion into slices perpendicular to the lowering direction. Each slice was characterised by SEM, EDX and X-ray powder diffraction (Cu Kα, two-circle diffractometer, Siemens D500).

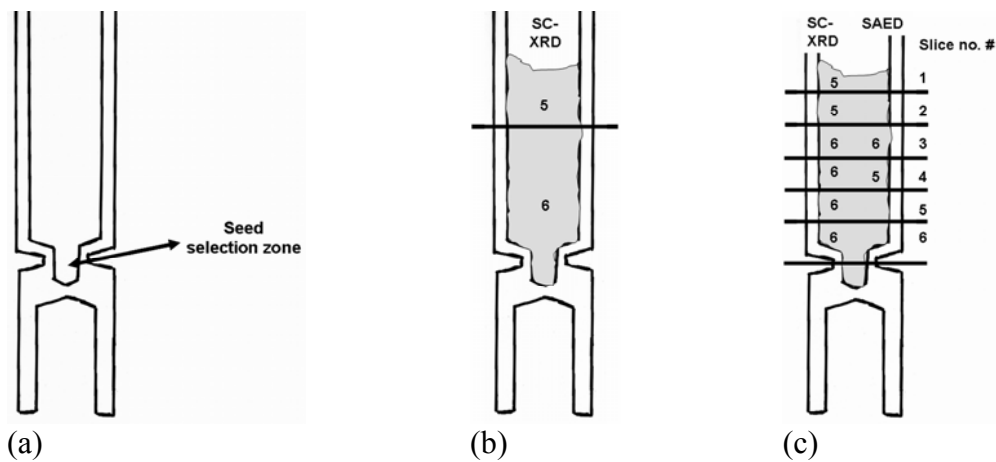

(a) (b) (c)

Fig. 1a (left), 1b (middle) and 1c (right):
(1a) schematic crucible design
(1b) cut crucible (sample 4); the numbers indicate the kind of superstructure found by SC-XRD: 5 means a 5×5×4.5 superstructure, 6 means a 6×6×5.5 superstructure
(1c) cut crucible sample 5; comparison of superstructures analysed by SAED and SC-XRD in the different slices; both sample 4 and sample 5 with starting composition 80.78 at% Cu.

For SAED (Jeol: JEM-4000 FX Electron Microscope, 400kV) small parts of selected slices were ground and applied on a copper grid using an ethanol dispersion.

For SC-XRD smaller parts of all slices with approximate volumes of $(50\text{-}100\mu m)^3$ were measured using an Oxford Diffraction diffractometer (Onyx CCD detector, Mo-K$\alpha$ radiation, graphite monochromator).

Results and discussion

DSC measurements of the Yb-Cu system for (80.3 to 79.4 at% Cu) are shown in Fig. 2a and 2b.

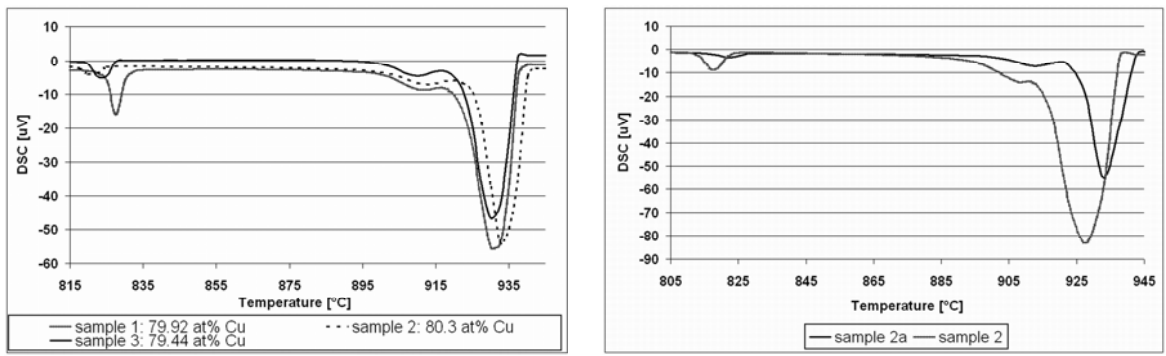

Fig. 2a (left) and 2b (right):
(2a) DSC measurements of Yb-Cu compounds of different compositions
(2b) The same material at different cooling rates is shown. Note that a shoulder can be seen in all DSC signals.

Fig. 2a shows three DSC curves for different Yb-Cu compositions. The Yb-Cu ratio was varied systematically: Yb:Cu = 1:3.86, 1:3.98, 1:4.08. The cooling rate for all three measurements was 4K/min. Fig. 2b compares two different cooling rates of sample 2 (10K/min and 4K/min).

Comparing with existing phase diagrams [1,2] (see also Fig. 8a) the cooling peak at about 830°C is clearly related to the peritectic formation of $YbCu_{3.5}$ at 825°C and the peak at about 935°C refers to the solidification of $YbCu_{4.5}$. Note that between these two events our DSC data clearly indicate another singularity of the heat capacity. The temperature difference between this singularity and the last solidification peak at 935°C remains constant when the cooling rate is varied (Fig. 2b).

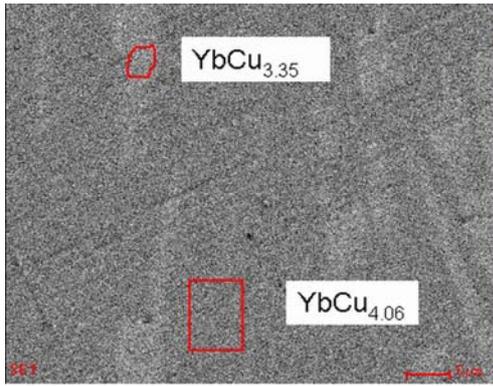

Fig. 3: SEM photograph of sample 3: two phase region, the corresponding compositions (EDX) are given

Fig. 3 shows a two-phase region (composition from EDX: $YbCu_{3.35}$ and $YbCu_{4.06}$) in a typical DSC run. No third phase, which would be expected from the DSC signal at about 910°C, can be detected. Nevertheless, this signal is reproducible as has been demonstrated by various experiments at different cooling rates (Fig. 2b). The idea is that there has to be some phase with either physical properties very similar to those of $YbCu_{4.5}$ or that some properitectic phase is completely transformed. Such phases may exist following Černý's building sequence [4].

To test this hypothesis, a Bridgman run was performed (initial composition 80.79 at% Cu is equivalent to a ratio Yb:Cu = 1:4.21). After growth the crucible was cut into six slices (slice #6 contains the first solidified part of the sample, slice #1 the last one, see Fig. 1c). All slices were checked by EDX. They are single phase, the compositions range between 80.35 at% and 79.6 at% Cu. Parts of slice #3 and #4 are analysed by SAED.

Fig. 4 shows the electron diffraction patterns of slice #3. On the basis of reciprocal lattice planes $h\ k\ 0$ and $h\ -h\ l$ the lattice constants of the superstructure $a^S = 41(1)$Å, $b^S = 41.9(7)$Å, and $c^S = 38.6(4)$Å and the monoclinic angle $\beta = 90.5(5)°$ can be identified. The data are displayed in Table 1.

| | | [3] | this work | | [5] |
|---|---|---|---|---|---|
| Ideal composition | at% Cu | 81.84 | 81.48 | 80.95 | 77.78 |
| | $x$ in $Cu_xYb$ | 4.5 | 4.4 | 4.25 | 3.5 |
| EDX composition | at% Cu | – | 80(1) | 80(1) | 77(1) |
| $n$ in $\{(Cu_5Yb)_n(Cu_2Yb)\}$ | | 5 | 4 | 3 | 1 |
| Superstructure $(n+2)\times(n+2)\times(n+1.5)$ | | 7×7×6.5 | 6×6×5.5 | 5×5×4.5 | 3×3×2.5 |
| $a^S$/Å | | 48.961(4) | 41(1) | 35.6(4) | |
| $b^S$/Å | | 48.994(4) | 41.9(7) | 35(1) | |
| $c^S$/Å | | 45.643(1) | 38.6(4) | 32.27(9) | |
| $\beta$/° | | 91.24(1) | 90.5(5) | 90.0(5) | |
| $A = a^S$/Å/$(n+2)$ | | | 6.9(4) | 7.021(9) | |
| $B = b^S$/Å/$(n+2)$ | | | 6.99(2) | 7.00(3) | |
| $C = c^S$/Å/$(n+1.5)$ | | | 6.9(1) | 7.04(2) | |

**Table 1.** Structural parameters for copper-rich phases $YbCu_x$ following the building principle $\{(YbCu_5)_n \cdot (YbCu_2)\}$. Lattice parameters (this work) were determined from SAED or SC-XRD.

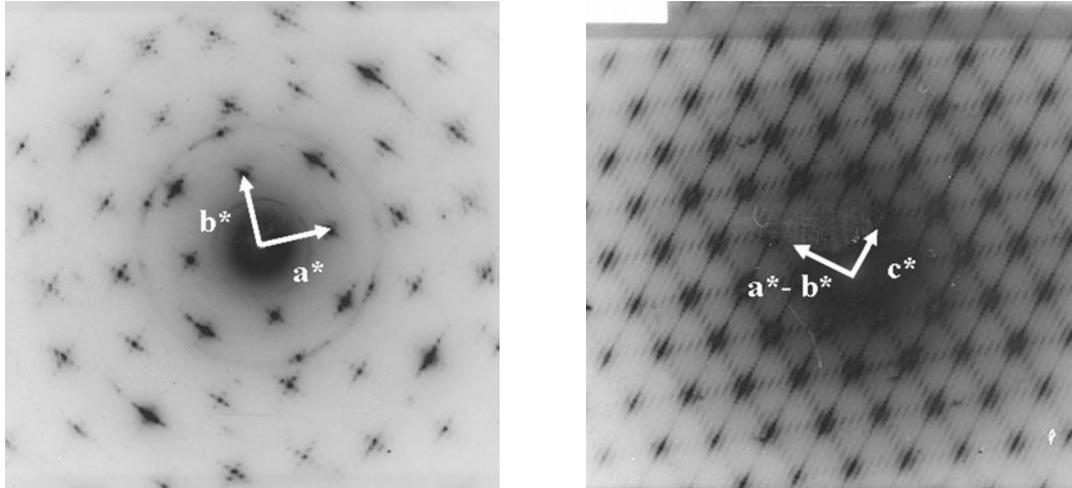

Fig. 4a (left) and 4b (right): Electron diffraction of the 6×6×5.5 superstructure (slice #3)
(4a) reciprocal lattice planes $h\,k\,0$ and
(4b) reciprocal lattice planes $h\,-h\,l$

Fig. 5 displays the [001] SAED pattern of slice #4 corresponding to the $h\,k\,0$ plane. From this plane one can identify $a^S = 35.6(4)$Å and $b^S = 35(1)$Å. No perpendicular direction could be oriented in the transmission electron microscope, but the corresponding information is contained in SC-XRD results (Fig. 6). Parts of all slices were checked by SC-XRD, thereby assigned superstructures are displayed in Fig. 1c.

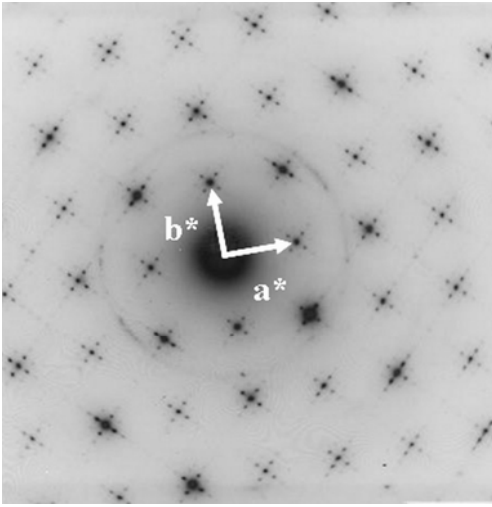

Fig. 5: Electron diffraction of the 5×5×4.5 superstructure (slice #4); reciprocal lattice plane $h\,k\,0$

In slice #1 and #2 a 5×5×4.5 superstructure is found. The remaining parts (slices #3 to #6) show a 6×6×5.5 superstructure. At first glance these results seem to be contradictory to the SAED results, where a 6×6×5.5 superstructure was found in slice #3 and a 5×5 superstructure in the a*b* plane in slice #4 (Fig. 1c). But in consideration of the only slight differences in composition between these phases, a competing growth behaviour cannot be excluded. The missing information about the c* direction of the 5×5×4.5 superstructure was gained from another Bridgman run (Fig. 1b). In the upper part SC-XRD reveals a 5×5×4.5 superstructure, in the lower one a 6×6×5.5 superstructure (Fig. 1b). The photographs in Fig. 6 are taken from the upper part of this run. The reciprocal lattice plane $h\,k\,0$ is analogue to that found by the electron diffraction (Fig. 5). The plane $h\,0\,l$ provides $c^S = 32.27(9)$Å and $\beta = 90.0(5)°$. For a better comparison the data are also listed in Table 1.

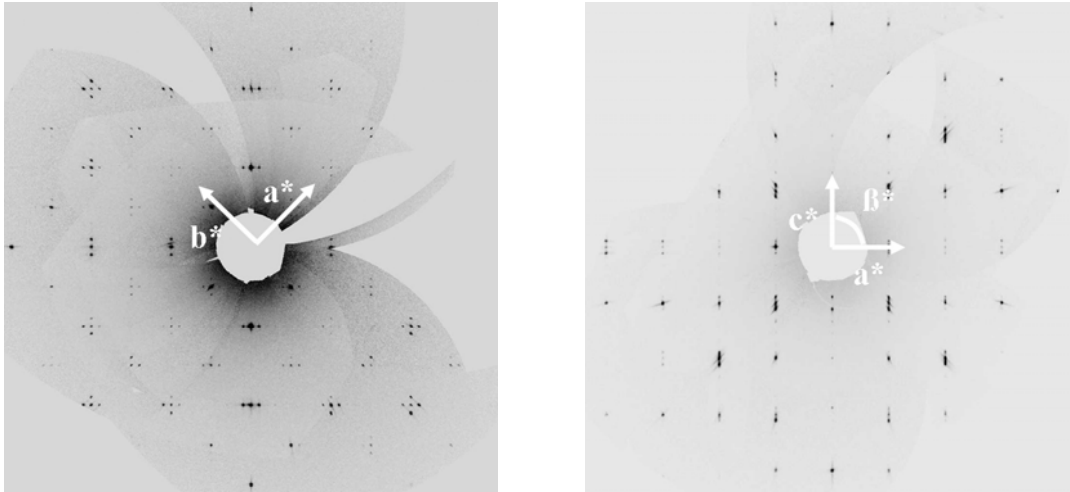

Fig. 6: Single crystal X-ray diffraction patterns of a 5×5×4.5 superstructure;
(6a) *h k* 0 plane and
(6b) *h* 0 *l* plane

Satellite reflections in all photographs are clearly visible. Their intensity modulation is comparable to that found for Dy-Cu phases [4]. According to the building principle $\{(RECu_5)_n \cdot (RECu_2)\}$ [4] the number of superstructure reflections corresponds to the two different phases $YbCu_{4.4}$ ($n = 4$) and $YbCu_{4.25}$ ($n = 3$), although the EDX results cannot resolve these two phases and the composition measured is about 80±1 at% Cu. To avoid misunderstandings one is speaking of $YbCu_{4.4}$ and $YbCu_{4.25}$ in the following. To learn more about the formation of these two new phases, DSC measurements with parts of slice #3 and #4 were done. Cooling rates from 1000°C to 700°C were chosen as 2K/min. The results can be seen in Fig. 7. Both curves show typical shoulders like in Fig. 2. The shoulder in $YbCu_{4.25}$ is not as pronounced as in the case of $YbCu_{4.4}$ but clearly the peak does not show the expected exponential decrease.

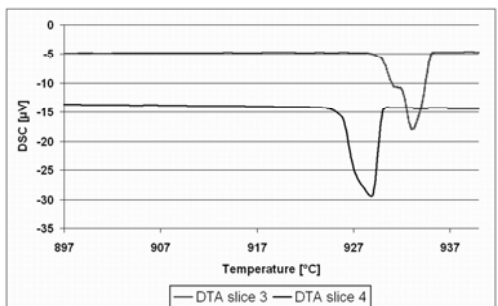

Fig. 7: DSC measurements of $YbCu_{4.4}$ (lower curve slice #4) and $YbCu_{4.25}$ (upper curve slice #3)

These shoulders can be contributed to peritectic reactions at (934±2)°C and (931±3)°C, respectively. These temperatures include baseline corrections.

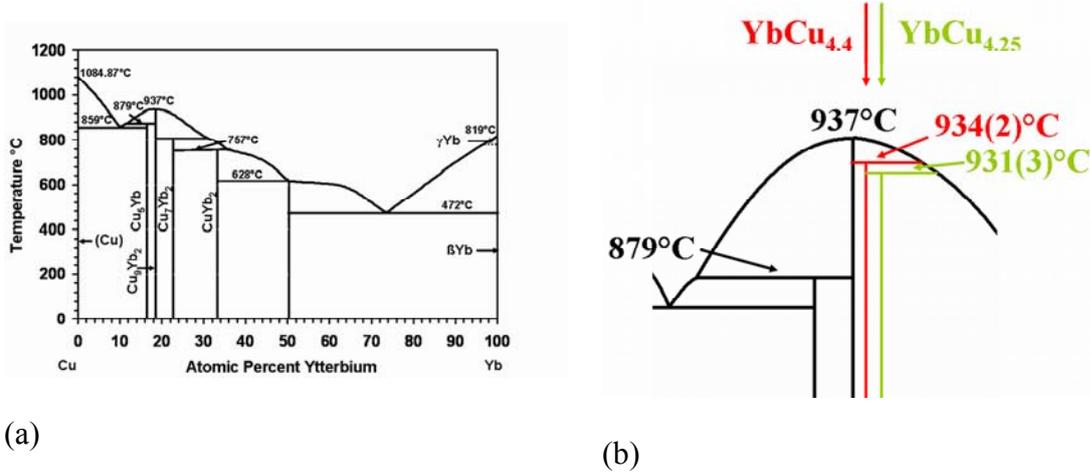

(a)

(b)

Fig. 8:
(8a) Anterior phase diagram of Yb-Cu (after [2])
(8b) modifications in the Cu-rich region (this work)

According to our results the Yb-Cu phase diagram in the copper rich region ($3.5 < x < 4.5$ in $YbCu_x$) is proposed to be modified as given in Fig. 8b. The phase width of $YbCu_{4.5}$ which is drawn from 80.5 to 81.8 at% Cu in the latest revision of the Yb-Cu phase diagram [5] splits into discrete lines at 81.8, 81.5 and 81.0 at% Cu corresponding to $YbCu_{4.5}$, $YbCu_{4.4}$ and $YbCu_{4.25}$, respectively.

Including this work the Černý phases $\{(RECu_5)_n \cdot (RECu_2)\}$ with $n = 1, 2, 3, 4$ and $5$ have been detected without gap experimentally now (Table 2). Interestingly, the end members are not all observed so far: $n = 0$ would mean a pure $MgCu_2$ (cF24) type structure but at this composition $YbCu_2$ crystallising in the $CeCu_2$ type (oI12) is realized instead [1]. For $n = \infty$, the pure phase has been reported for $RE$ = Yb not at ambient conditions but as a high pressure phase only [10]. Thus the end members of the series seem not to exist but rather their intergrowth variants. A still open question remains the behaviour of the $RE$-Cu systems for $n \geq 6$. We already have some hints pointing to the existence of larger superstructures. More detailed investigations may clarify this point in the future.

| $n$ | at% Cu | Proved for $RE$ = | Remark | Reference |
|---|---|---|---|---|
| 0 | 66.67 | – | no MgCu$_2$-type (cF24) for any $RE$; CeCu$_2$-type (oI12) for $RE$ = Yb, Dy, Er, Ho, Gd | [1] |
| 1 | 77.78 | Dy | $RE_2$Cu$_7$ composition reported in phase diagrams for $RE$ = Lu, Tm, Er, Ho, Tb, Gd (high temperature phase?) | [1], [4] |
|   |   |   | Yb (isostructural?) | [5] |
| 2 | 80.00 | Dy |   | [4] |
| 3 | 80.95 | Yb |   | (this work) |
| 4 | 81.48 | Yb |   | (this work) |
| 5 | 81.84 | Yb, Dy | $RE_2$Cu$_9$ composition reported in phase diagrams for $RE$ = Lu, Tm, Er, Ho, Tb, Gd (isostructural?) | [1], [3], [4] |
| 6 | 82.05 | – | $\Delta$/at% = 0.21; refers to ($n$ - 1) |   |
| 7 | 82.22 | – | $\Delta$/at% = 0.17 |   |
| 8 | 82.35 | – | $\Delta$/at% = 0.13 |   |
| 9 | 82.46 | – | $\Delta$/at% = 0.11 |   |
| 10 | 82.54 | – | $\Delta$/at% = 0.08 |   |
| ... |   |   |   |   |
| ∞ | 83.33 | Yb | AuBe$_5$-type (cF24) reported for $RE$ = Lu, Tm, Er; high pressure phase for Yb; low temperature phase for $RE$ = Dy, Ho, Tb, Gd | [1], [10] |

**Table 2.** Predicted and proved Černý phases {($RE$Cu$_5$)$_n$·($RE$Cu$_2$)} in $RE$-Cu systems.

Summary

Due to careful structural and thermodynamic investigations in the copper-rich region in the Yb-Cu system we obtained evidence of two new phases, namely YbCu$_{4.4}$ and YbCu$_{4.25}$. These two phases are formed peritectically at 934(2)°C and 931(3)°C, respectively, and are part of a series of monoclinic superstructures based on the AuBe$_5$ structure type. YbCu$_{4.4}$ represents a 6×6×5.5 and YbCu$_{4.25}$ a 5×5×4.5 superstructure.

Both phases have been predicted by Černý`s formula but this work verifies their existence for the first time experimentally. Unlike in the Dy-Cu system the monoclinic distortion is less pronounced (β ≈ 90°) and thus tetragonal symmetry cannot been ruled out. These results shed more light on the complex situation in the Cu-rich region of the Yb-Cu phase diagram. From a crystal grower's point of view it is a sophisticated challenge to grow mm-seized single crystalline material of a deliberately chosen superstructure as compositions and formation temperatures are very close.

Acknowledgements

This work was carried out in the framework of the European Network of Excellence "Complex Metallic Alloys": The absorption of travel costs connected to our Long Term Cu-*RE* project is gratefully scknowledged.